\documentstyle[12pt]{article}
\begin{document}

\hfill UB-ECM-PF 93/22

\hfill December 1993

\vspace*{3mm}

\begin{center}

{\LARGE \bf
Renormalization-group improved effective potential for
interacting
theories with several mass scales in curved spacetime}

\vspace{4mm}

\renewcommand
\baselinestretch{0.8}
\medskip

{\sc E. Elizalde}
\footnote{E-mail: eli @ ebubecm1.bitnet} \\
Department E.C.M. and I.F.A.E., Faculty of Physics,
University of  Barcelona, \\ Diagonal 647, 08028 Barcelona, \\
and Blanes Center for Advanced Studies, C.S.I.C., 17300 Blanes,
Spain \\
 and \\
{\sc S.D. Odintsov}\footnote{E-mail: odintsov @ ebubecm1.bitnet.
On
leave from: Tomsk Pedagogical Institute, 634041 Tomsk, Russian
Federation.} \\ Department E.C.M., Faculty of Physics,
University of  Barcelona, \\  Diagonal 647, 08028 Barcelona,
Spain \\

\renewcommand
\baselinestretch{1.4}

\vspace{5mm}

{\bf Abstract}

\end{center}

The renormalization group (RG) is used in order to obtain the RG
improved effective potential in curved spacetime. This potential
is
explicitly calculated for the Yukawa model and for scalar
electrodynamics, i.e. theories with several (namely, more than
one)
mass scales, in a space of constant curvature. Using the $\lambda
\varphi^4$-theory on a general curved spacetime as an example,
we
show how it is possible to find the RG improved effective
Lagrangian in curved spacetime. As specific applications, we
discuss the possibility of curvature induced phase transitions in
the
Yukawa model
and the effective equations (back-reaction problem) for the
$\lambda \varphi^4$-theory on a De Sitter background.

\vspace{4mm}

\newpage

\section{Introduction}

It is commonly accepted nowadays that the early universe
experienced an
inflationary epoch \cite{1,2} during its evolution (for an
introduction
and a review of
different models of inflationary universe, see \cite{3}-\cite{6}).
 Some of the models of inflationary universes are based
on the
effective potential of the scalar field. As a rule, a flat space
effective potential is used in such models for the description
of
the
effective cosmological constant. However, such an approach cannot
be
completely consistent, since the curvature was not so small at
this
epoch.

Thus, the necessity of calculating the effective potential
(or, more generally, the effective action \cite{7}) in curved
spacetime
appears naturally motivated by cosmological considerations.
However, the
problem of obtaining the effective action in interacting
theories in curved spacetime can currently be solved only in
part,
owing
to the fact that even one-loop calculations can be done just on
some
very specific backgrounds and for quite simple interacting
theories
(see, for instance, \cite{8}-\cite{14}; for a general review and an
extensive
list of references see \cite{8}). (Note that the effective action
can be
also found as an expansion on terms of fixed order of the
curvature
tensors; however, such technique can be applied only in spaces
with
small, slowly varying curvature.) It is  quite an old idea
\cite{15}
that of using a kind of
effective action formalism for investigating the
quantum-corrected Einstein equations (in Refs. \cite{15} this
approach
has been discussed for conformally invariant scalar fields mainly).
However,
this program is not so easy to realize.

Given this situation, where even the one-loop effective action in
general
curved spacetime is hard to find, some indirect methods may prove
of
great importance. In particular, the flat space effective
potential
\cite{16,17} can be found in the leading (or even subleading)
logarithmic approximation using the renormalization group (RG)
equations. Such RG improved effective potential \cite{16,18,19}
yields
the sum of the leading logarithms of the whole perturbation
theory
and
actually provides the extension for the range of the scalar
field,
if we
compare it with the one-loop effective potential. Recently, the
RG
improved effective potential for massive theories in flat
spacetime
has
been discussed \cite{20}-\cite{22}. It has been found in these
references
that
such effective potential (which may be important for the
discussion
of
stability of the electroweak vacuum in the standard model)
acquires
several new features when compared with the one for the massless
theory
\cite{16}.

First of all, the effective cosmological constant (or running
vacuum
energy) which is obtained from the transformation of the trivial
$V(0)$
into a $\Phi$-dependent quantity by the RG, appears already in
flat
space
\cite{20}-\cite{22}. As we will see below, a most consistent and
natural
interpretation of this effective cosmological constant among the
other
running couplings is obtained in curved space, where it is not
necessary
to introduce any artificial functions \cite{20} which make the
effective
potential satisfy the RG equations.

Second, in theories which have several mass scales there appears
a
problem connected with the choice of the RG parameter, which is
not
unique there. Some solution of this problem ---based on the use of
the
decoupling theorem \cite{23} and of the effective field theory
\cite{24}---
has been given in Ref. \cite{22} using the Yukawa model as an
example.

The purpose of the present work is to study the RG improvement
of
the
effective potential (or, more generally, of the effective
Lagrangian) in
massive theories in curved spacetime. The method of Ref.
\cite{22}
can
be extended to curved spacetime and provides a rigorous treatment
for
the RG improved effective potential when there are a few
effective
mass
scales in the theory. (Notice that, generally speaking, the RG
in
curved
spacetime has been constructed in Refs. \cite{25}-\cite{27} (see
\cite{8}
for a
review), but this version of the RG is convenient for the study
of
the
high-energy asymptotics mainly and not in the context under
discussion here).

The paper is organized as follows. In the next section the RG
improved
effective potential is obtained for the Yukawa model in curved
spacetime for the case of constant curvature. In section 3 we
present
the same calculation using scalar electrodynamics as an example.
The RG
improved effective Lagrangian for the $\lambda \Phi^4$-theory in
a
general, slowly varying spacetime is found in section 4. In
section
5 we
investigate the curvature-induced phase transitions for the
Yukawa
model
in the region where scalar particles are very heavy. Section 6 is
devoted to the discussion of the effective equations on a De Sitter
background, using the $\lambda \varphi^4$-theory as an example.
Finally,
Sect. 7 includes conclusions and some additional remarks. \bigskip

\section{The RG improved effective potential for the Yukawa model
in
curved spacetime}

Let us consider the following Lagrangian for the Yukawa model in
curved spacetime
\begin{equation}
L_m = \frac{1}{2} g^{\mu\nu} \partial_\mu \varphi  \partial_\nu
\varphi - \frac{1}{2} m^2 \varphi^2 + \frac{1}{2} \xi R \varphi^2
- \frac{\lambda}{4!} \varphi^4 + \bar{\psi} \left( i \gamma^\mu
(x)
\nabla_\mu - h \varphi \right) \psi,
\label{1}
\end{equation}
where $\psi$ is a massless, $N$-component Dirac spinor and
$\varphi$ is a massive scalar.

In what follows we will restrict ourselves to spacetimes of
constant curvature, and we will find the RG improved effective
potential for the theory (\ref{1}). As usually, in order to
obtain
from (\ref{1}) a multiplicatively renormalizable theory, one has
to
add to (\ref{1}) the action of the external gravitational field,
 i.e.
\begin{equation}
L_{ext} = \Lambda + \kappa R +a_1 R^2 +a_2
C_{\mu\nu\alpha\beta}^2
+a_3 G,
\label{2}
\end{equation}
where  $C_{\mu\nu\alpha\beta}$ is the Weyl tensor and $G$ is the
Gauss-Bonnet invariant. Note that the $\Box R$-term is absent
from
(\ref{2}), because of the restriction to constant curvature
spaces.

Now, since the theory is multiplicatively renormalizable, its
effective potential satisfies the standard RG equation:
\begin{equation}
DV \equiv \left( \mu \frac{\partial}{\partial \mu} + \beta_i
\frac{\partial}{\partial \lambda_i} - \gamma \Phi
\frac{\partial}{\partial \Phi} \right) V (\mu,\lambda_i,
\Phi)=0,
\label{3}
\end{equation}
where $\lambda_i = (\xi, \lambda, h^2,m^2, \Lambda, \kappa, a_1,
a_2,a_3)$, $\beta_{\lambda_i}$ are the corresponding
$\beta$-functions, and $\Phi$ is the background scalar. The only
difference  with respect to the analogue of (\ref{3}) in flat
space
is that, among the constants $\lambda_i$ we have some
gravitational coupling
constants which do not appear in flat space. (In this case
$L_{ext}
=\Lambda$.)

The solution of Eq. (\ref{3}) by the method of the characteristics
gives
\begin{equation}
V  (\mu,\lambda_i,\Phi)= V (\mu e^t,\lambda_i (t),\Phi
(t)),
\label{4}
\end{equation}
where
\begin{eqnarray}
\frac{d \lambda_i (t)}{dt} = \beta_i (\lambda_i (t)), &&
\lambda_i (0) = \lambda_i, \nonumber \\
\frac{d \Phi (t)}{dt} =-\gamma (t) \Phi (t), &&
\Phi (0) = \Phi.
\label{5}
\end{eqnarray}
As usually, the physical meaning of (\ref{4}) is that the
effective
potential is found provided its functional form at some certain
$t$
is known.

The one-loop $\beta$-functions of the theory can be calculated
using the standard background field method. Of course, the
$\beta$-functions corresponding to $\lambda$, $h$, $m^2$,
$\Lambda$
and $\Phi$ have the same form as in  flat space. Let us write all
$\beta$-functions explicitly in the notations of Ref. \cite{22}
which were introduced for flat space
\begin{eqnarray}
&& \beta_\lambda = \frac{1}{(4\pi)^2} (3\lambda^2 +8N\lambda
h^2-48
Nh^4) \equiv \beta_{\lambda 1} \lambda^2 +\beta_{\lambda h}
\lambda
h^2+\beta_{\lambda hh} h^4, \nonumber \\  && \beta_h =
\frac{1}{(4\pi)^2}
(4N+6) h^4 \equiv \beta_{h1} h^4, \nonumber \\ &&  \beta_{m^2}
= \frac{1}{(4\pi)^2} (\lambda +4N h^2) m^2 \equiv
(\gamma_{m\lambda
1} \lambda + \gamma_{mh1} h^2) m^2, \ \ \  \gamma =
\frac{2Nh^2}{(4\pi)^2} = \gamma_1 h^2, \nonumber \\  &&  \beta_\xi
= (\gamma_{m\lambda 1} \lambda + \gamma_{mh1} h^2) (\xi - 1/6),
\
\ \ \beta_\Lambda = \frac{m^4}{2(4\pi)^2}, \ \ \   \beta_{\kappa}
= \frac{m^2 (\xi -1/6)}{(4\pi)^2},  \nonumber \\ &&  \beta_{a_1} =
\frac{(\xi - 1/6)^2}{2(4\pi)^2},   \ \ \  \beta_{a_2} =
\frac{1}{120(4\pi)^2} (1+6N),
\ \ \  \beta_{a_3} =- \frac{1}{360(4\pi)^2} (1+11N).
\label{5p}
\end{eqnarray}
The solution of the one-loop RG equations for all coupling
constants can be found in the form
\begin{eqnarray}
&& h^2(t) = \frac{h^2}{1- \beta_{h1} h^2 t}\equiv \frac{h^2}{B(t)},
\ \
\ \ \Phi (t) = \Phi B(t)^{\gamma_1/\beta_{h1}}, \nonumber
\\ &&\lambda (t) = h^2 \frac{a (\lambda -bh^2) B(t)^{
a \beta_{\lambda1} /\beta_{h1}-1} -b (\lambda -ah^2)
B(t)^{b \beta_{\lambda1} /\beta_{h1}-1}}{ (\lambda
-
bh^2) B(t)^{a \beta_{\lambda1} /\beta_{h1}} -
(\lambda -ah^2) B(t)^{b \beta_{\lambda1}
/\beta_{h1}}}, \nonumber \\ && m^2(t) = m^2 B(t)^{-
\gamma_{ mh1} /\beta_{h1}} \left( \frac{ (\lambda -bh^2)
B(t)^{a \beta_{\lambda1} /\beta_{h1}} - (\lambda
-ah^2)
B(t)^{b \beta_{\lambda1} /\beta_{h1}}}{(a-b)h^2}
\right)^{-\gamma_{m\lambda 1} /\beta_{h1}} , \nonumber \\
&&\xi (t) = \frac{1}{6}+ \left( \xi - \frac{1}{6} \right)
B(t)^{-\gamma_{ mh1} /\beta_{h1}} \left( \frac{
(\lambda -bh^2) B(t)^{a \beta_{\lambda1}
/\beta_{h1}} - (\lambda -ah^2) B(t)^{b
\beta_{\lambda1} /\beta_{h1}}}{(a-b)h^2} \right)^{-\gamma_{m
\lambda 1} /\beta_{h1}} , \nonumber \\  && a_2(t) = a_2 + \frac{t
(1+6N)}{120(4\pi)^2},
\ \ \  a_3(t) = a_3- \frac{t (1+11N)}{360(4\pi)^2}, \nonumber \\ &&
 \Lambda (t) =
\Lambda + \frac{m^4}{2(4\pi)^2} \int_0^t   dt \,
B(t)^{-2\gamma_{ mh1} /\beta_{h1}} \nonumber \end{eqnarray}
\[ \left( \frac{
(\lambda -bh^2) B(t)^{a \beta_{\lambda1}
/\beta_{h1}} - (\lambda -ah^2) B(t)^{b
\beta_{\lambda1} /\beta_{h1}}}{(a-b)h^2} \right)^{-
2\gamma_{m\lambda 1} /\beta_{h1}}
\equiv
\Lambda + \frac{m^4}{2(4\pi)^2} A(t), \]
\begin{equation} \kappa (t) = \kappa +\frac{m^2}{(4\pi)^2}
\left( \xi -
\frac{1}{6}\right) A(t),
\ \ \  a_1 (t) = a_1 +\frac{1}{2(4\pi)^2} \left( \xi -
\frac{1}{6} \right)^2  A(t).
\label{6}
\end{equation}
Here $a$ and $b$ are the roots of the equation
\[
\beta_{\lambda 1} y^2 + ( \beta_{\lambda h} - \beta_{h1} )y +
\beta_{\lambda hh} =0 \]
(see also \cite{22}).

Now the question appears of what is the choice of the RG
parameter
$t$ which leads to the summation of all the logarithms to all
orders. Working in the one-loop approximation we have two
logarithms appearing in the one-loop effective potential, namely
\begin{equation}
\log \frac{ \frac{1}{2} \lambda \Phi^2 +m^2 - (\xi -1/6)
R}{\mu^2},
\ \ \ \ \log \frac{h^2 \Phi^2 - \frac{1}{4} R}{\mu^2}.
\label{7}
\end{equation}
(Notice that in the problem under discussion we consider the
effective potential as an expansion on the curvature invariants.
Hence, $\lambda \Phi^2$ and $h^2 \Phi^2$ are dominant when
compared
with the $R$-terms in (\ref{7}). Note also that in massless
theories the choice of $t$ is actually unique: $t=\frac{1}{2}
\log
(\Phi^2 /\mu^2)$ \cite{28}.)

Hence, we have two effective masses in the theory, and there
seems
no way open to a choice of $t$ which eliminates the leading
logarithms to all orders ---as a consequence of the existence of
these two different mass scales. However, there exists some
solution to this problem, that has been suggested in flat space
\cite{22}. Such solution is based on the decoupling theorem
\cite{23} and in the effective field theory \cite{24}. It can
actually
be
generalized to curved spacetime, as we will see below.

Indeed, let us consider the region $h^2\Phi^2 \geq m^2$. There,
\begin{eqnarray}
\log  \frac{ \frac{1}{2} \lambda \Phi^2 +m^2 - (\xi -1/6)
R}{\mu^2}
&\simeq & \log   \frac{h^2 \Phi^2- \frac{1}{4} R}{\mu^2} + \log
\frac{ \frac{1}{2} \lambda \Phi^2 +m^2 - (\xi -1/6) R}{h^2\Phi^2}
\nonumber \\
&\simeq & \log   \frac{h^2 \Phi^2- \frac{1}{4} R}{\mu^2} + \log
\left( \frac{\lambda}{2h^2} + \frac{m^2}{h^2\Phi^2} - \frac{ (\xi
-
1/6) R}{h^2\Phi^2} \right).
\label{8}
\end{eqnarray}
One can see that the second term on the rhs in (\ref{8}) is small
if compared with the first one (note that $h^2 >\lambda$), i.e.,
in
the region $h^2\Phi^2 \geq m^2$ we effectively have just a single
mass scale which is the natural one to be used as the RG
parameter
$t$.

The standard procedure, in which we employ the tree-level
effective
potential as boundary function, gives us the following leading log
approximation for the effective potential in the region
$h^2\Phi^2
\geq m^2$:
\begin{equation}
V= \frac{1}{2} m^2(t) \Phi^2 (t) - \frac{1}{2} \xi (t) \Phi^2 (t)
R + \frac{1}{4!} \lambda (t) \Phi^4 (t) - \Lambda (t) - \kappa (t)
R -
a_1 (t)
R^2 - a_2 (t) C_{\mu\nu\alpha\beta}^2 - a_3 (t) G,
\label{9}
\end{equation}
where the effective couplings  in (\ref{9}) are given by Eqs.
(\ref{6}) and
\begin{equation}
t= \frac{1}{2} \log   \frac{h^2 \Phi^2- \frac{1}{4} R}{\mu^2}.
\end{equation}
Thus, we have been able to construct the RG improved effective
potential for a quite complicated Yukawa theory in curved
spacetime (for a discussion of the RG improved potential in curved
spacetime in the massless case, see \cite{28}). This potential
performs
the summation of all the
leading logarithms in perturbation theory and, in this sense, it is
much
more exact than the one-loop effective potential, because it
takes
into account all orders of the perturbation theory.  Notice that
even the calculation of the one-loop effective potential for the
theory (\ref{1}) in curved spacetime is very hard to do. From our
result (\ref{8}) we can  get the one-loop effective potential
very
easily as an expansion of (\ref{8}) when $|t| << 1$ (with
suitable
renormalization conditions, which must be imposed after
expanding).
This ends the analysis of the region $h^2\Phi^2 \geq m^2$.

In order to investigate the region $m^2 \geq h^2\Phi^2$, we adopt
the method developed in Ref. \cite{22} for the same theory in flat
space. According to this method, the scalar particle $\Phi$ is
treated as a very heavy particle of mass $m$. In such a
formulation, the effects of a heavy particle lead to a shift of
the
parameters of the low-energy theory. We will choose only
logarithmic terms to be responsible for this shift, because in
a
direct calculation of the one-loop effective potential the
non-logarithmic terms depend on the scheme (or, in other words,
on
the choice of renormalization conditions). The parameters of the
low-energy theory are defined by
\begin{eqnarray}
&& \widetilde{\lambda} = \lambda + \frac{3\lambda^2}{2(4\pi)^2}
\log
\frac{m^2}{\mu^2}, \ \ \ \widetilde{\Phi} = \Phi, \ \ \
\widetilde{m}^2 = m^2 + \frac{\lambda m^2}{2(4\pi)^2} \log
\frac{m^2}{\mu^2}, \nonumber \\  && \widetilde{\kappa} = \kappa +
\frac{m^2(\xi
-1/6)}{2(4\pi)^2} \log \frac{m^2}{\mu^2}, \ \ \
\widetilde{\xi} = \xi + \frac{\lambda (\xi - 1/6)}{2(4\pi)^2} \log
\frac{m^2}{\mu^2},  \ \ \
\widetilde{\Lambda} = \Lambda + \frac{m^4}{4(4\pi)^2} \log
\frac{m^2}{\mu^2}, \nonumber \\
&& \widetilde{a}_1 = a_1 + \frac{(\xi -1/6)^2}{4(4\pi)^2} \log
\frac{m^2}{\mu^2}, \ \ \ \widetilde{a}_2 = a_2 +
\frac{1}{240(4\pi)^2}
\log \frac{m^2}{\mu^2}, \ \ \ \widetilde{a}_3 = a_3 -
\frac{1}{720(4\pi)^2} \log \frac{m^2}{\mu^2}. \nonumber \\
\label{11}
\end{eqnarray}
Notice that $h^2$ gets also shifted, but from two-loop
corrections
\cite{22}: \\ $\widetilde{h}^2=h^2 + [3h^4/(4\pi)^2] \log
(m^2/\mu^2)$.

Using Eqs. (\ref{11}) we may transform the RG operator acting on
the effective potential (\ref{3}) as follows
\begin{equation}
D = (D\mu) \frac{\partial}{\partial \mu} + (D\widetilde{\lambda}_i)
\frac{\partial}{\partial \widetilde{\lambda}_i} +
(D\widetilde{\Phi})
\frac{\partial}{\partial \widetilde{\Phi}} =  \mu
\frac{\partial}{\partial
\mu} + \widetilde{\beta}_{\lambda_i} \frac{\partial}{\partial
\widetilde{\lambda}_i} - \widetilde{\gamma} \widetilde{\Phi}
 \frac{\partial}{\partial
\widetilde{\Phi}},
\end{equation}
where from (\ref{11}) and (\ref{5}) it follows that (in the
one-loop approximation)
\begin{eqnarray}
&& \widetilde{\beta}_\lambda = D \widetilde{\lambda} =
\frac{1}{(4\pi)^2}
(8N\lambda h^2-48Nh^4), \ \ \  \widetilde{\beta}_h = D
\widetilde{h}^2 =
\frac{4Nh^4}{(4\pi)^2}, \nonumber \\
&&   \widetilde{\beta}_{m^2} = D \widetilde{m}^2 =
\frac{4Nh^2}{(4\pi)^2} m^2, \
\ \
\gamma =\widetilde{\gamma}, \ \ \ \widetilde{\beta}_\xi = D
\widetilde{\xi} =
\frac{4Nh^2}{(4\pi)^2} \left( \xi - \frac{1}{6} \right), \nonumber
\\
&& \widetilde{\beta}_\Lambda = \widetilde{\beta}_\kappa =
\widetilde{\beta}_{a_1} =0,
\
\ \  \widetilde{\beta}_{a_2} =  \frac{6N}{120(4\pi)^2}, \  \ \
\widetilde{\beta}_{a_3} = - \frac{11N}{360(4\pi)^2}.
\end{eqnarray}
Solving the RG equations for the effective couplings with tilde,
we
find
\begin{eqnarray}
&& \widetilde{h}^2 (t)= \widetilde{h}^2 \left( 1 - \frac{4N
\widetilde{h}^2}{(4\pi)^2}
t \right)^{-1}, \ \ \ \widetilde{\Phi} (t) = \widetilde{\Phi}
\left( 1 -
\frac{4N
\widetilde{h}^2}{(4\pi)^2} t \right)^{1/2}, \ \ \
\widetilde{\beta}_{h1}
=\frac{4N}{(4\pi)^2}, \nonumber \\
&& \widetilde{\lambda} (t)= \left( \widetilde{\lambda}-
\frac{\beta_{\lambda
hh}}{\widetilde{\beta}_{h1}- \beta_{\lambda h}}
\widetilde{h}^2 \right) \left(
1
- \frac{4N \widetilde{h}^2}{(4\pi)^2} t \right)^{-\beta_{\lambda
h}/\widetilde{\beta}_{h1}}+\frac{\beta_{\lambda hh}\widetilde{h}^2
(t)}{
\widetilde{\beta}_{h1}- \beta_{\lambda h}}, \nonumber \\
&& \widetilde{m}^2 (t)= \widetilde{m}^2 \left( 1 - \frac{4N
\widetilde{h}^2}{(4\pi)^2}
t \right), \ \ \ \widetilde{\Lambda} (t) = \widetilde{\Lambda},  \
\ \
\widetilde{\kappa} (t) = \widetilde{\kappa}, \ \ \ \widetilde{\xi}
(t) = \frac{1}{6}
+
\left(\widetilde{\xi}- \frac{1}{6} \right) \left( 1 - \frac{4N
\widetilde{h}^2}{(4\pi)^2} t \right), \nonumber \\
&& \widetilde{a}_1 (t) = a_1, \ \ \ \widetilde{a}_2 (t) = a_2 +
\frac{6Nt}{120(4\pi)^2}, \ \ \  \widetilde{a}_3 (t) = a_3 -
\frac{11Nt}{360(4\pi)^2}.
\label{14}
\end{eqnarray}

Summing all leading logarithms, the RG improved effective
potential
is now given by
\begin{eqnarray}
V &=& \frac{1}{2} \widetilde{m}^2 (t) \widetilde{\Phi}^2 (t)
-\frac{1}{2}
\widetilde{\xi} (t) \widetilde{\Phi}^2 (t) R + \frac{1}{4!}
\widetilde{\lambda} (t)
\widetilde{\Phi}^4 (t) - \widetilde{\Lambda} (t) -
\widetilde{\kappa} (t)R - \widetilde{a}_1
(t) R^2 \nonumber \\
&& - \widetilde{a}_2 (t) C_{\mu\nu\alpha\beta}^2 -  \widetilde{a}_3
(t) G, \
\ \
\ \ \ t= \frac{1}{2} \log \frac{\widetilde{h}^2\widetilde{\Phi}^2
- R/4}{\mu^2}.
\label{15}
\end{eqnarray}
It is interesting to note that the effective conformal coupling
$\widetilde{\xi} (t)$  grows with $t$ as in the case of the
asymptotically free
models of Refs. \cite{8,26,27}.

Agreement between the potentials (\ref{9}) and (\ref{15}) can be
easily checked \cite{22}. Indeed, both of them satisfy the same
RG
equation and are independent on $\mu$. Choosing $\mu = m$ and
considering the region $g\Phi \simeq \widetilde{g} \widetilde{\Phi}
=m$, one
sees
that $t\simeq 0$ both in (\ref{9}) and in (\ref{15}), and hence the
effective potentials reduce to the same boundary function (tree
potential) and here coincide.

To summarize, in this section we have generalized the approach of
Ref. \cite{22} to curved spacetime, and found the RG improved
effective potential for the Yukawa theory.
\bigskip

\section{RG improved potential for scalar electrodynamics in
curved
spacetime}

We now consider the simple model of an abelian gauge theory
---scalar electrodynamics--- in curved spacetime. We are going
to
show how the method developed in the previous section can be
easily
applied to gauge theories, considering scalar electrodynamics as
an
example. In other words, we will find the RG improved effective
potential for scalar electrodynamics in a curved spacetime of
constant curvature.

The classical Lagrangian for the theory is given by
\begin{equation}
L=L_m + L_{ext},
\end{equation}
where
\begin{equation}
L_m = \frac{1}{2} \left( \partial_\mu \Phi_1- e A_\mu \Phi_2
\right)^2 + \frac{1}{2} \left( \partial_\mu \Phi_2+ e A_\mu
\Phi_1
\right)^2 -\frac{1}{2} m^2 \Phi^2 + \frac{1}{2} \xi R  \Phi^2 -
\frac{1}{4!} \lambda \Phi^4 - \frac{1}{4} F_{\mu\nu} F^{\mu\nu},
\end{equation}
$L_{ext}$ is given by (\ref{2}), and $\Phi^2 = \Phi_i  \Phi_i
=
\Phi_1^2 + \Phi_2^2$.

We will adopt the Landau gauge, in which the one-loop
$\beta$-function for the gauge parameter is zero, and the RG
equation for the scalar effective potential has the same form
(\ref{3}) as in the Yukawa theory (except for the change $h^2
\rightarrow e^2$ in $\lambda_i$). The solution of the RG equation
(\ref{3}) has the same formal structure (\ref{4}), being now the
one-loop $\beta$-functions given by
\begin{eqnarray}
&& \beta_\lambda = \frac{1}{(4\pi)^2} \left( \frac{10}{3}
\lambda^2
-12 e^2 \lambda +36 e^4 \right), \ \ \ \beta_{e^2} =
\frac{2e^4}{3(4\pi)^2}, \ \ \ \beta_{m^2} =  \frac{m^2}{(4\pi)^2}
\left( \frac{4}{3} \lambda - 6e^2 \right), \nonumber \\
&& \gamma = - \frac{3e^2}{(4\pi)^2}, \ \ \ \beta_\xi =
\frac{(\xi
- 1/6)}{(4\pi)^2} \left( \frac{4}{3} \lambda - 6e^2 \right), \
\ \
\beta_\Lambda =  \frac{m^4}{(4\pi)^2}, \ \ \ \beta_\kappa =
\frac{2m^2(\xi - 1/6)}{(4\pi)^2}, \nonumber \\
&& \beta_{a_1} = \frac{(\xi - 1/6)^2}{(4\pi)^2}, \ \ \
\beta_{a_2}
= \frac{7}{60(4\pi)^2}, \ \ \ a_3 (t) = a_3- \frac{8t}{45(4\pi)^2}.
\end{eqnarray}
The solutions of the RG equations for the coupling constants are
\begin{eqnarray}
&& e^2(t) = e^2 \left( 1- \frac{2e^2t}{3(4\pi)^2} \right)^{-1},
\
\ \ \Phi^2 (t) = \Phi^2 \left( 1- \frac{2e^2t}{3(4\pi)^2}
\right)^{-9}, \nonumber \\
&& \lambda (t) = \frac{1}{10} e^2(t) \left[ \sqrt{719} \, \tan
\left(  \frac{1}{2} \sqrt{719} \, \log e^2(t) +C \right) +19
\right], \nonumber \\
&& C = \arctan  \left[ \frac{1}{\sqrt{719} } \left( \frac{10
\lambda}{e^2} -19 \right) \right] - \frac{1}{2} \sqrt{719} \,
\log
e^2, \nonumber \\
&& m^2(t)= m^2 \left[ \frac{e^2(t)}{e^2} \right]^{-26/5}
\frac{\cos^{2/5} \left( \frac{1}{2} \sqrt{719} \, \log e^2 +C
\right) }{\cos^{2/5} \left( \frac{1}{2} \sqrt{719} \, \log e^2(t)
+C \right) }, \nonumber \\
&& \xi (t) = \frac{1}{6} + \left(\xi - \frac{1}{6} \right)
 \left[ \frac{e^2(t)}{e^2} \right]^{-26/5} \frac{\cos^{2/5}
\left(
\frac{1}{2} \sqrt{719} \, \log e^2 +C \right) }{\cos^{2/5} \left(
\frac{1}{2} \sqrt{719} \, \log e^2(t) +C \right) }, \nonumber \\
&& a_2(t) = a_2 + \frac{7 t}{60(4\pi)^2}, \ \ \ a_3(t) =a_3-
\frac{8t}{45(4\pi)^2}, \nonumber \\
&& \lambda (t) = \Lambda + m^4 A_1 (t), \ \ \ \kappa (t) = \kappa
+ 2m^2 (\xi -1/6) A_1(t), \ \ \ a_1(t) = a_1 + (\xi -1/6)^2
A_1(t),
\nonumber \\
&& A_1(t) = \int_0^t \frac{dt}{(4\pi)^2} \,  \left[
\frac{e^2(t)}{e^2} \right]^{-52/5} \frac{\cos^{4/5} \left(
\frac{1}{2} \sqrt{719} \, \log e^2 +C \right) }{\cos^{4/5} \left(
\frac{1}{2} \sqrt{719} \, \log e^2(t) +C \right) }.
\label{19}
\end{eqnarray}
Notice that the flat-space coupling constants (\ref{19}) have
been
obtained in the classical work \cite{16}. We see again that there
naturally appears the effective cosmological constant $\Lambda
(t)$, which is essential even in flat space \cite{20,21,22}, as
a
RG improved vacuum energy (effective potentials at zero
background
fields). In curved spacetime we have a very consistent
interpretation of $\Lambda (t)$ as one of the coupling constants
for the $L_{ext}$ sector, and there is no need to introduce special
ways of getting $\Lambda (t)$ in some artificial form \cite{20}.

Now the question of the choice of the RG parameter $t$ appears.
Here the situation is even more complicated, as compared with the
case of the Yukawa model, because after diagonalization of the
mass
matrix we get the following effective masses in the theory:
\begin{equation}
m_1^2 = m^2 + \frac{\lambda}{2} \Phi^2 - (\xi - 1/6)R, \ \ \
m_2^2
= m^2 + \frac{\lambda}{6} \Phi^2 - (\xi - 1/6)R, \ \ \ m_{\mu\nu}^2
= \delta_{\mu\nu} (e^2\Phi^2 -R/4).
\end{equation}
Using the same techniques as in the previous section, we may
consider first the region $e^2\Phi^2> m^2 + (\lambda/2) \Phi^2$. In
this
region the natural choice for the RG parameter $t$ is
\begin{equation}
t= \frac{1}{2} \log \frac{e^2\Phi^2 -R/4}{\mu^2}.
\label{21}
\end{equation}
Substituting (\ref{19}), with $t$ as in (\ref{21}), into the RG
improved effective potential defined by the tree potential, we
have
the same form (\ref{9}) for the RG improved potential.

In the region $m^2 >e^2\Phi^2$, where the scalar particles are
very
heavy, we may again define the low-energy theory parameters.
Then,
as in the previous section, the coupling constants corresponding
to
the effective theory can be cast into the form
\begin{eqnarray}
&& \widetilde{e}^2(t) = \widetilde{e}^2 \left( 1-
\frac{2\widetilde{e}^2 t}{3(4\pi)^2}
\right)^{-1}, \ \ \ \widetilde{\Phi}^2(t) = \widetilde{\Phi}^2
\left( 1 -
\frac{2\widetilde{e}^2 t}{3(4\pi)^2} \right)^{-9}, \ \ \
\widetilde{m}^2(t) =
\widetilde{m}^2 \left( 1- \frac{2\widetilde{e}^2 t}{3(4\pi)^2}
\right)^{9}, \nonumber
\\
&& \widetilde{\xi} (t) = \frac{1}{6} + \left(\widetilde{\xi} -
\frac{1}{6}
\right) \left( 1- \frac{2\widetilde{e}^2 t}{3(4\pi)^2} \right)^{9},
\ \
\
\widetilde{\Lambda} (t) = \widetilde{\Lambda}, \ \ \
\widetilde{\kappa} (t) =
\widetilde{\kappa}, \nonumber \\ && \widetilde{\lambda} (t) =
\frac{54}{19}
\widetilde{e}^2(t)
+
\left[ \left( \lambda -  \frac{54}{19}  \widetilde{e}^2
\right)^{-1/19}
\frac{\widetilde{e}^2(t)}{\widetilde{e}^2} \right]^{-19}, \nonumber
\\
&& \widetilde{a}_1 (t) = \widetilde{a}_1, \ \ \   \widetilde{a}_2
(t) = \widetilde{a}_2+
\frac{t}{10(4\pi)^2}, \ \ \   \widetilde{a}_3 (t) =
\widetilde{a}_3-
\frac{31t}{180(4\pi)^2}.
\label{22}
\end{eqnarray}
As we see, the conformal effective coupling constant does not
show
asymptotically a conformally invariant behavior, as happened in
the
Yukawa model above.

The RG improved effective potential is given by
\begin{eqnarray}
V &=& \frac{1}{2} \widetilde{m}^2 (t) \widetilde{\Phi}^2 (t)
-\frac{1}{2}
\widetilde{\xi} (t) \widetilde{\Phi}^2 (t) R + \frac{1}{4!}
\widetilde{\lambda} (t)
\widetilde{\Phi}^4 (t) - \widetilde{\Lambda} (t) -
\widetilde{\kappa} (t)R - \widetilde{a}_1
(t) R^2 \nonumber \\
&& - \widetilde{a}_2 (t) C_{\mu\nu\alpha\beta}^2 -  \widetilde{a}_3
(t) G, \
\ \
\ \ \ t= \frac{1}{2} \log \frac{\widetilde{e}^2\widetilde{\Phi}^2
-
R/4}{\mu^2},
\label{23}
\end{eqnarray}
with the effective couplings being given by (\ref{22}).

Thus, we have been able to construct the RG improved effective
potential for scalar electrodynamics. Proceeding in a similar way,
one can in principle construct it for more complicated theories
(as
GUTs), where more effective mass scales are present.

Another important point concerns now the possibility to
generalize
the above results to non-constant curvature spaces. In the next
section we will demonstrate that this is indeed feasible.
\bigskip

\section{RG improved effective Lagrangian in the
$\lambda \Phi^4$-theory in curved spacetime}

Let us consider the  $\lambda \Phi^4$-theory with the Lagrangian
\begin{equation}
L=L_m + L_{ext}, \ \ \ \ L_m= \frac{1}{2} g^{\mu\nu} \partial_\mu
\varphi \partial_\nu \varphi - \frac{1}{2} (m^2 - \xi R)
\varphi^2 - \frac{\lambda}{24} \varphi^4 +a_5 \Box \varphi^2,
\end{equation}
in general curved spacetime. Now curvature will not be constant,
in general, but we shall make the assumption that the
gravitational field is slowly varying. In this case, $L_{ext}$ as
given by (\ref{2}) contains also the term $a_4 \Box R$ which is
absent when the curvature is constant. Notice that we have also
added to $L_m$ the total derivative term $a_5 \Box \varphi^2$
which becomes important at the quantum level, as we will see
below.

Using the same technique as in Sect. 2 and the explicit form of
the one-loop $\beta$-functions of the   $\lambda \Phi^4$-theory
(notice however that here the coupling constants $a_4$ and $a_5$
are to be considered together with the other coupling constants of
the
theory), we get the RG improved effective Lagrangian under the
following form
\begin{eqnarray}
L_{eff} &=&  \frac{1}{2} g^{\mu\nu} \partial_\mu \Phi
\partial_\nu \Phi - \frac{1}{2} [m^2(t) - \xi (t) R] \Phi^2 -
\frac{\lambda (t)}{24} \Phi^4 +a_5 (t) \Box \Phi^2
+ \Lambda (t) \nonumber \\
&&+ \kappa (t) R + a_1(t) R^2 +a_2 (t) C_{\mu\nu\alpha\beta}^2
+a_3(t) G +a_4(t) \Box R. \label{25}
\end{eqnarray}
Here \[ t= \frac{1}{2} \log \frac{m^2- (\xi - 1/6) R +\lambda
\Phi^2/2}{\mu^2},\]
and
\begin{eqnarray}
&&\lambda (t) = \lambda \left( 1- \frac{3\lambda t}{(4\pi)^2}
\right)^{-1}, \ \ \ m^2(t) = m^2  \left( 1- \frac{3\lambda
t}{(4\pi)^2} \right)^{-1/3}, \nonumber \\
&&\xi (t) = \frac{1}{6}+ \left( \xi - \frac{1}{6} \right)  \left(
1- \frac{3\lambda t}{(4\pi)^2} \right)^{-1/3}, \ \ \ \Lambda (t)
=
\Lambda - \frac{m^4}{2\lambda} \left( 1- \frac{3\lambda
t}{(4\pi)^2} \right)^{1/3} + \frac{m^4}{2\lambda}, \nonumber \\
&&\kappa (t) = \kappa -\frac{m^2}{\lambda} \left( \xi -
\frac{1}{6}
\right) \left[ \left( 1- \frac{3\lambda t}{(4\pi)^2}
\right)^{1/3}
-1
\right],
\nonumber \\ && a_1 (t) = a_1 -\frac{1}{2\lambda} \left( \xi -
\frac{1}{6}
\right)^2 \left[ \left( 1- \frac{3\lambda t}{(4\pi)^2}
\right)^{1/3}
-1\right],
\nonumber \\ && a_2 (t) = a_2 +\frac{t}{120 (4\pi)^2}, \ \ \  a_3
(t)
=
a_3
-
\frac{t}{360(4\pi)^2}, \nonumber \\ && a_4 (t) = a_4 +\frac{t}{180
(4\pi)^2} +\frac{1}{3\lambda} \left( \xi - \frac{1}{6} \right)
\left[ \left( 1- \frac{3\lambda t}{(4\pi)^2} \right)^{2/3}
-1\right],
\nonumber \\ && a_5 (t) = a_5 +\frac{1}{36} \ln  \left( 1-
\frac{3\lambda
t}{(4\pi)^2} \right).
\end{eqnarray}
However, the RG improved effective Lagrangian gives now some
additional information, because $\Phi$ and $R$ in (\ref{25})
are not constant but may depend on spacetime coordinates.

In the same way, we can construct the RG improved effective
Lagrangian for more complicated theories, like the Yukawa model,
scalar electrodynamics, or some other theories. The approach
developed above still works in those cases and the only
supplementary difficulties that appear are of technical nature.
Such difficulties are connected with the more involved structure
of the RG parameter $t$ ---owing to the fact that we must then
diagonalize a more involved effective mass matrix with
non-constant
(i.e., coordinate dependent) elements. Notice also the
new terms of type $\Box R$ and $\Box \Phi^2$.

For the sake of conciseness, we are not going to discuss in the
present case the RG improved effective Lagrangian for theories
with several mass scales. This effective Lagrangian will be
however extremely useful in the discussion of dynamical, quantum
corrected Einstein equations (for an introduction, see
\cite{15}), which are very difficult to solve due to the high
non-linearity of the problem. In what follows we will restrict
ourselves to the discussion of some cosmological applications of
the  RG improved effective potential above.
\bigskip

\section{Curvature induced phase transitions in the Yukawa model
in curved spacetime}

It is generally accepted nowadays that the very early universe
experienced several phase transitions before it could reach its
present state. It is also very possible that one of those phase
transitions could be induced by the external gravitational field
existing at an early epoch \cite{10,12}. We will here investigate
this possibility using as main ingredient the RG improved
effective potential for the case of the Yukawa model. For
simplicity, we will just consider the region $m^2 \geq h^2
\Phi^2$ and restrict ourselves to the linear curvature
approximation. We shall also put
$\widetilde{\Lambda}=\widetilde{\kappa}=0$,
since this just amounts to a  simple rescaling of the vacuum
energy.
Implementing these conditions in the RG improved effective
potential (\ref{14}), we get
\begin{eqnarray}
V&=& \frac{1}{2} \widetilde{m}^2 \left( 1 - \frac{4N
\widetilde{h}^2}{(4\pi)^2}
t \right)^{2} \widetilde{\Phi}^2 - \frac{1}{2} \left[ \frac{1}{6}
+
\left( \widetilde{\xi} - \frac{1}{6} \right)  \left( 1 - \frac{4N
\widetilde{h}^2}{(4\pi)^2}
t \right)\right] \widetilde{\Phi}^2
\left( 1 - \frac{4N \widetilde{h}^2}{(4\pi)^2} t \right)R \nonumber
\\ &&
+  \frac{1}{4!} \widetilde{\Phi}^4  \left[ \widetilde{\lambda} - 12
\widetilde{h}^2   +12 \widetilde{h}^2  \left( 1 - \frac{4N
\widetilde{h}^2}{(4\pi)^2}
t \right) \right],
\label{poti}
\end{eqnarray}
with \[ t= \frac{1}{2} \log \frac{\widetilde{h}^2
\widetilde{\Phi}^2-
R/4}{\mu^2}.\]

We can now start with the analysis of critical points corresponding
to
the potential (\ref{poti}) both for the massless and for the
massive
case. Calling in general $x \equiv \widetilde{\Phi}^2/\mu^2$ and $y
\equiv
R/\mu^2$, so that $V = V(x,y)$,  let us remember that the critical
 parameters, $x_c$, $y_c$, corresponding to the first-order
phase transition are found from the conditions
\begin{equation}
V(x_c,y_c)=0, \ \ \ \ \left. \frac{\partial V}{\partial x}
\right|_{x_c,y_c} =0, \ \ \ \ \left. \frac{\partial^2 V}{\partial
x^2} \right|_{x_c,y_c} >0.
\label{cp}
\end{equation}
For the RG improved potential
they lead to some transcendental equations which cannot be solved
analytically. However, we do will obtain some particular solutions.
We shall be concerned with first-order phase transitions where the
order parameter $\widetilde{\Phi}^2$ experiences a quick change for
some
critical value, $R_c$, of the curvature.

The first two equations (\ref{cp}) for the potential (\ref{poti})
are (after some immediate recombination and having eliminated
already
the trivial solution at (0,0)): \begin{eqnarray}
 2\widetilde{m}^2 C(t) -R  \left[ \frac{1}{6} + \left(
\widetilde{\xi} - \frac{1}{6}
\right) C(t) \right] + \frac{x}{12}  \left[ (\widetilde{\lambda} -
12
\widetilde{h}^2 ) C(t)^{-1} +12 \widetilde{h}^2  \right] &=&0,
\nonumber \\
 -2\widetilde{m}^2 C(t) +R  \left[ \frac{1}{6} +2
\left( \widetilde{\xi} - \frac{1}{6} \right) C(t) \right] +
\frac{(4\pi)^2x}{24 N\widetilde{h}^2}
\left[ \widetilde{\lambda} - 12 \widetilde{h}^2  +12
\widetilde{h}^2 C(t) \right]
-\widetilde{h}^2 x &=&0, \nonumber \\
 C(t) \equiv 1 - \frac{4N \widetilde{h}^2}{(4\pi)^2} t. &&
\label{eq2}
\end{eqnarray}
 Two particular choices of $\xi$ are interesting \cite{sbb}, in
each of
the following two cases. \medskip

\noindent{\it (a) Massless case}: $\widetilde{m}^2 =0$, with
$\widetilde{h}^2 <1$ and
$\widetilde{\lambda}<1$.

We have carried out two different analysis,
 corresponding to $\widetilde{\xi} =1/6$ and $\widetilde{\xi}
=10^4$,
respectively. The result in both cases is the same: Eqs.
(\ref{eq2}) are
incompatible, so that the only critical point is obtained at
$\widetilde{\Phi}_c^2=0$,  $R_c=0$.
\medskip

\noindent{\it (b) Massive case}: $\widetilde{m}^2 \neq 0$, with
$\widetilde{h}^2 <1$
and $\widetilde{\lambda}<1$.

Now the result changes completely, a non-trivial critical point
exists, and its location and properties depend clearly on the
value of  $\widetilde{\xi}$ considered. For $\widetilde{\xi} =1/6$,
we obtain
\begin{equation}
\widetilde{\Phi}^2_c\simeq
\frac{4!N\widetilde{h}^2\widetilde{m}^2}{(4\pi)^2} \left(
\frac{N\widetilde{h}^4}{(4\pi)^2}- \lambda \right)^{-1} \simeq
\frac{48
\widetilde{m}^2}{\widetilde{h}^2}, \ \ \ \ \ R_c \simeq
12\widetilde{m}^2. \end{equation}
On the other hand, for  $\widetilde{\xi} =10^4$, the result is
\begin{equation}
\widetilde{\Phi}^2_c\simeq \frac{4!4
N\widetilde{h}^2\widetilde{m}^2}{(4\pi)^2}
\left(
\frac{N\widetilde{h}^4}{(4\pi)^2}- \widetilde{\lambda} \right)^{-1}
\simeq \frac{96
\xi
\widetilde{m}^2}{\widetilde{h}^2}, \ \ \ \ \ R_c \simeq
\frac{2\widetilde{m}^2}{\widetilde{\xi}}.
\end{equation}
Apart from the additional factor of 2 in the expression for
$\widetilde{\Phi}_c^2$, notice the specific behavior of $R_c$ in
terms of
$\widetilde{\xi}$ obtained from the last case.
We observe that for small values of $\widetilde{\lambda}$, namely
$\widetilde{\lambda}
< N\widetilde{h}^4 /(4\pi )^2$,  an admissible, positive solution
is
obtained
in
this massive case. Moreover, it is immediate to check the
consistency of
such solution of the quite involved, transcendental equations
(\ref{eq2}). In fact, it is obtained for $t$ of the order of $1$,
a
condition that is easily matched {\it a posteriori}:
\[
t \sim \frac{1}{2} \log \frac{[48- 1/(4\widetilde{\xi})]
\widetilde{m}^2}{\mu^2} \sim
1. \]
We finish this section by pointing out that what we have found
above is
a very interesting example of curvature-induced phase transition
which
has the property that it takes place only at non-zero mass and it
is
absent in the massless case.
 \bigskip

\section{Effective equations in De Sitter space}

Let us discuss now the effective equations for the $\lambda
\varphi^4$-theory in a De Sitter background. This may be viewed
as an attempt at studying the qualitative features of quantum
corrected gravitational equations in a static geometry (the back
reaction problem).

It is rather natural to start from some curved background in this
problem, since flat spacetime is unstable as a background
\cite{30}.
However, it is well known \cite{15} that this problem is very
difficult
to solve technically, due to the appearence of higher derivatives,
of
its high non-linearity, etc. Recently some interesting approach has
been
discussed \cite{31} that reduces the problem to second-order
equations.
This technique may prove to be useful also in our context, however,
notice that  we have here many more non-linearities,
connected
with the several logarithms.

Our Lagrangian is chosen in the form (in this section
Euclidean notations will be used)
\begin{equation}
L= \frac{1}{2} \left( \partial_\mu \varphi \right)^2 +
\frac{1}{2} m^2 \varphi^2  + \frac{\lambda}{4!} \varphi^4  +
\frac{1}{2} \xi R \varphi^2 + \Lambda_0 + \kappa R +
\widetilde{a}_1
R^2.
\label{61}
\end{equation}
(Notice that $G=R^2/6$ in such a background, and hence only the
$R^2$ term is induced.) We have put a subindex to $\Lambda$ in
(\ref{61}).

Repeating the same procedure of Sect. 4, with some minor sign
modifications, we can easily get the RG improved effective
Lagrangian on the De Sitter background
\begin{equation}
L_{eff} = \frac{1}{2} m^2(t) \Phi^2
+ \frac{\lambda (t)}{4!} \varphi^4  + \frac{1}{2} \xi (t) R
\varphi^2 + \Lambda_0(t) + \kappa (t) R + \widetilde{a}_1 (t) R^2,
\label{62}
\end{equation}
where
\[
 \lambda (t) = \lambda \left( 1- \frac{3\lambda t}{(4\pi)^2}
\right)^{-1}, \  m^2 (t) = m^2 \left( 1- \frac{3\lambda
t}{(4\pi)^2} \right)^{-1/3}, \
 \xi (t) = \frac{1}{6} + \left( \xi - \frac{1}{6} \right)
\left( 1- \frac{3\lambda t}{(4\pi)^2} \right)^{-1/3}, \] \[
 \Lambda_0 (t) = \Lambda_0 -  \frac{m^4}{2\lambda}  \left( 1-
\frac{3\lambda t}{(4\pi)^2} \right)^{-1/3}+
\frac{m^4}{2\lambda}, \ \ \
 \kappa (t) = \kappa - \frac{m^2}{\lambda}  \left( \xi -
\frac{1}{6} \right) \left[  \left( 1- \frac{3\lambda t}{(4\pi)^2}
\right)^{1/3} -1 \right], \]
\begin{equation}  \widetilde{a}_1 (t) =
 \widetilde{a}_1 - \frac{1}{2\lambda}  \left( \xi - \frac{1}{6}
\right)
\left[  \left( 1- \frac{3\lambda t}{(4\pi)^2} \right)^{1/3} -1
\right]-\frac{t}{2160 (4\pi)^2},
\end{equation}
and being
\[ t= \frac{1}{2} \log \frac{m^2+ (\xi - 1/6) R +\lambda
\Phi^2/2}{\mu^2}.\]
Using the properties of the De Sitter space:
\begin{equation}
R=4\Lambda, \ \ \ \ \int d^4x \, \sqrt{g} =
\frac{24\pi^2}{\Lambda^2},
\end{equation}
we may study the effective equations along the line of Ref.
\cite{14} (see also \cite{35}). First of all, let us introduce the
variables
$y=1/\Lambda$ and $x=\Phi^2/\Lambda$. Then the effective action
is given by
\begin{equation}
\frac{S_{eff}}{24\pi^2} = \frac{1}{2} m^2(t) xy + \frac{\lambda
(t)}{4!} x^2 + 2\xi (t) x + \Lambda_0 (t) y^2 + 4 \kappa (t) y
+ 16 \widetilde{a}_1 (t),
\end{equation}
with \[ t= \frac{1}{2} \log \frac{m^2y+4 (\xi - 1/6) +\lambda
x/2}{\mu^2 y}.\]

For simplicity, let us consider only the massless case: $m^2=0$.
Presumably, the introduction of mass will not change the results
of the following qualitative analysis. Then,
\begin{equation}
\frac{S_{eff}}{24\pi^2} = \frac{\lambda (t)}{4!} x^2 + 2\xi (t)
x + \Lambda_0  y^2 + 4 \kappa  y + 16 \widetilde{a}_1 (t).
\end{equation}
The variational equations obtained from this effective action are
\begin{eqnarray}
&& \frac{\lambda (t) x}{12} + 2 \xi (t) + \frac{B(x)}{2 (4\pi)^2}
\left( x + \frac{8 (\xi - 1/6)}{\lambda} \right)^{-1} =0 ,
\nonumber \\
&&  2y \Lambda_0 + 4 \kappa - \frac{B(x)}{2 (4\pi)^2y}
+\Phi^2\left[ \frac{\lambda (t) x}{12} + 2 \xi (t) +\frac{B(x)}{2
(4\pi)^2}  \left( x + \frac{8 (\xi - 1/6)}{\lambda}
\right)^{-1} \right] =0, \label{sys} \end{eqnarray}
where
\[ B(x) = \frac{3\lambda^2x^2}{4!} F(t)^{-2} + 2\lambda x \left(
\xi - \frac{1}{6} \right) F(t)^{-4/3}
+ 8 \left( \xi - \frac{1}{6} \right)^2 F(t)^{-2/3} - \frac{1}{135},
\ \ \ \ F(t) \equiv 1- \frac{3\lambda t}{(4\pi)^2}. \]

For $\Phi =0$, $\Lambda_0 \neq 0$, then as in Ref. \cite{14} we
get a perturbative solution of the kind:
\[
\Lambda \simeq -\frac{\Lambda_0}{2\kappa} + {\cal O} (\hbar), \
\ \ \kappa <0.
\]
If $\Lambda_0=0$, we get the same quantum solution as in Ref.
\cite{32}, namely, for $\xi =1/6$
\begin{equation}  \Lambda = - 1080  (4\pi)^2 \kappa. \label{108}
\end{equation}
For $\xi \neq 1/6$ we get iterative corrections to (\ref{108}).

For $\Phi \neq 0$ the system (\ref{sys}) should be analyzed more
carefully.
We get two solutions of (\ref{sys}), perturbatively in $\xi - 1/6$,
in
the following way (plus and minus signs go together)
\begin{eqnarray}
\Phi^2  &=& \frac{4\Lambda_0 F(t)^2}{\kappa \lambda \left[ F(t)
+\frac{3\lambda}{4(4\pi)^2}
\right]} \left\{ 1\pm \left[ 1 +  \frac{\Lambda_0
F(t)^2}{\kappa^2 2(4\pi)^2 \left[ F(t)
+\frac{3\lambda}{4(4\pi)^2} \right]^2} \right]^{1/2} \right\}^{-1},
\nonumber
\\
\Lambda  &=&- \frac{\Lambda_0}{\kappa} \left\{ 1\pm \left[ 1 +
\frac{\Lambda_0 F(t)^2}{\kappa^2 2(4\pi)^2 \left[
F(t) +\frac{3\lambda}{4(4\pi)^2} \right]^2}
\right]^{1/2} \right\}^{-1}.
\label{solius}
\end{eqnarray}
 Of course these are not explicit expressions, since $t$ involves
the
variable $\Phi^2$ itself (notice that solving the transcendental
equations
(\ref{sys}) explicitly is an impossible task). Eqs. (\ref{solius})
are to
be used recurrently, in the ordinary way: for $\lambda$ small and
 $\xi \neq 1/6$ ($\xi - 1/6$ small), starting from
\begin{equation}
t_0 =  \frac{1}{2} \log \frac{4(\xi - 1/6)\Lambda_0}{\mu^2}, \ \ \
\ \ \
\Phi_0
= \Phi(t_0), \ \ \ \ \ \ \Lambda_{(0)}=\Lambda (t_0),
\label{42}\end{equation}
one defines
\begin{equation}
t_1 =t_0 +  \frac{1}{2} \log \left( 1+\frac{\lambda\Phi_0^2}{2(\xi
-1/6)\Lambda_{(0)}}  \right), \ \ \ \ \ \ \Phi_1
= \Phi(t_1), \ \ \ \ \ \ \Lambda_1 =\Lambda (t_1), \end{equation}
and so on (do not confuse the $\Lambda_{(0)}$ here with the
$\Lambda_0$
above). In general,
\begin{equation}
t_{n} =t_0 +  \frac{1}{2} \log \left( 1+\frac{\lambda\Phi_{n-1}^2}{
2(\xi - 1/6)\Lambda_{n-1}} \right), \ \ \ \ \ \ \Phi_n =
\Phi (t_n), \ \ \ \ \ \ \Lambda_n =\Lambda (t_n). \end{equation}
It is interesting to see explicitly the results that one gets after
the first step of the iteration:
\begin{eqnarray}
\frac{1}{\Lambda} & \simeq & -
\frac{\kappa}{\Lambda_0} \left( 1 \pm \sqrt{1 +
\frac{\Lambda_0}{\kappa^2 2 (4\pi)^2}} \right), \nonumber \\
\Phi^2  & \simeq &  \frac{4\Lambda_0}{\kappa\lambda} \,
\left(1- \frac{3\lambda}{ 2 (4\pi)^2} \log \frac{4(\xi
-1/6)\Lambda}{\mu^2}\right) \left( 1 \pm \sqrt{1 +
\frac{\Lambda_0}{\kappa^2 2 (4\pi)^2}}\right)^{-1}.
\label{45}
 \end{eqnarray}
These expressions already provide a good approximation for the
assumed conditions.

The case when $m^2=0$ and simultaneously $\xi = 1/6$, is a special
one. It is certainly not obtained as a limiting case of (\ref{45}),
even
if this was obtained perturbatively in $\xi - 1/6$, since the
starting
point (\ref{42}) is now different. In fact, in this
situation, provided
\[ t =  \frac{1}{2} \log \left( \frac{\lambda \Phi^2}{2\mu^2}
\right)
\sim 1,
\]
---so that, as a consequence, $F(t) \sim 1$ (this may be viewed as
a
condition to define $\mu^2$)--- we also obtain
sensible expressions for the approximate solution to first order in
the recurrence, of the kind
\begin{eqnarray}
\frac{1}{\Lambda} & \simeq & -
\frac{\kappa}{\Lambda_0} \left( 1 \pm \sqrt{1 +
\frac{\Lambda_0}{\kappa^2 2 (4\pi)^2}} \right), \nonumber \\
\Phi^2  & \simeq &
\frac{4\Lambda_0}{\kappa\lambda} \left( 1 \pm \sqrt{1 +
\frac{\Lambda_0}{\kappa^2 2 (4\pi)^2}} \right)^{-1}.
\label{46}
\end{eqnarray}
We observe here a difference with respect to the situation when
$\Lambda_0
=0$ from the begining, Eq. (\ref{108}), that we will now comment
on.  As
a general observation,  we see that in all these cases we get a
reasonable solution
(with positive sign) which can be consistently given a perturbative
form, in the sense that quantum corrections adopt the form of small
corrections to the classical solutions (as it should be, in
accordance
with our approach). However, we must also point out that when we
have
a negative sign in (\ref{45}) or (\ref{46}) (which can be obtained
for
reasonable values of the parameters, too), we get a new quantum
solution
\[
\Lambda \simeq 4 (4\pi)^2\kappa,
\]
which has the same structure as in (\ref{108}). Actually, the
solution
(\ref{108}) is also obtained in a different limit, in which the
term
$-1/135$ in the definition of $B(x)$ becomes dominant, and it goes
down
in the square root of (\ref{46}) (as $1- \Lambda_0
/\kappa^2 540(4\pi)^2$).

 It is interesting
to note that for this quantum solution of the effective equations
the
cosmological constant $\Lambda_0$ is arbitrary but the curvature
does
not depend on it. Of course this solution is not realistic, as it
is
indicated by the transition that takes place to anti-De Sitter
space
($\kappa <0$, $\Lambda <0$), and our starting proposal was $\Lambda
>0$.

Summing up, we have shown explicitly that it is possible, in
principle, to apply the RG improved effective potential to the
back-reaction problem. It is clear, however, that already for the
case of a  static geometry the system of equations (\ref{sys})
coming from the effective potentials is quite involved.
\bigskip

\section{Conclusions}

In this paper we have constructed a RG improved effective
potential for theories which involve a few mass scales in curved
spacetime. Among the examples that have been discussed there are
the Yukawa model, scalar electrodynamics and the $\lambda
\varphi^4$-theory. This last example has been employed to show that
the method is very general and can be used to carry out the full
program of construction of a RG improved effective Lagrangian.
Moreover, the procedure can be extended also to the case of
quantum gravity.

Indeed, let us consider the theory of quantum $R^2$-gravity
with the following Lagrangian,
\begin{equation}
L = \Lambda  - \frac{1}{\kappa^2} \, R + a \left( R_{\mu\nu}^2 -
\frac{1}{3} R^2 \right)+ \frac{1}{3} b R^2.
\label{fex1}
\end{equation}
It is well known that this theory is multiplicatively
renormalizable and asymptotically free (for a general review, see
\cite{8}). The RG equations in a gauge of Landau type is given by
\cite{33}
\begin{eqnarray}
&&\frac{da}{dt} = \frac{13.3}{(4\pi)^2}, \ \ \  \frac{du}{dt} =-
\frac{1}{(4\pi)^2a}\left(  \frac{10}{3} u^2+\frac{183}{10} u+
\frac{5}{12} \right), \ \ \  \frac{d\kappa^2}{dt} =
\frac{\kappa^2}{(4\pi)^2a}\left(  \frac{10}{3} u-\frac{1}{3u}
\right), \nonumber \\
&& \frac{d\Lambda}{dt} = \frac{1}{(4\pi a)^2\kappa^4}\left(
\frac{5}{2} +\frac{1}{8u^2} \right)+ \frac{\Lambda}{2a}\left(
10+\frac{1}{u} \right), \ \ \ \ \ \ u \equiv - \frac{b}{a}.
\label{fex2}
\end{eqnarray}
Notice that, as should be the case, the RG equations for the
coupling
constants $a$ and $u$ are gauge independent, while those for
$\Lambda$ and $\kappa^2$ are gauge dependent. We expect that in a
gauge of Landau type (in which the RG equations for $\Lambda$ and
$\kappa^2$ are written) there is no one-loop $\beta$-function for
the gauge parameter in Eq. (\ref{3}), at least in the case of a
background of constant curvature.

Concerning now the choice of the RG parameter $t$ (and restricting
ourselves to a background of constant curvature), we observe that
we have a few different masses in our theory. We may consider the
(non-physical) region where, for instance, $\Lambda$ is dominant
as compared with $|R|/\kappa^2$. Then, naturally, $t=\frac{1}{2}
\log (\Lambda / \mu^2)$, and
\begin{equation}
L_{eff} = \Lambda (t) - \frac{1}{\kappa (t)} R + {\cal O} (R^2).
\label{fex3}
\end{equation}
Solving now Eqs. (\ref{fex2}) we obtain $L_{eff}$ explicitly. In a
similar way we can consider other regions. Our main purpose here
has been to show that, in fact, the above considerations can be
applied as well to the theory of quantum gravity itself. We shall
not
continue here the discussion of the properties of $R^2$-gravity
in any more detail since, as is known, such a theory is apparently
not consistent (an unitarity problem exists).

Note, finally, that in this paper we have limited ourselves to
investigate only one
application of the RG in curved spacetime to the effective
Lagrangian.
Some other applications, very well studied in the literature,
include the calculation of multiloop $\beta$-functions \cite{34},
asymptotics of vertex functions for strong curvature
\cite{8,25,26,27},
etc.

\vspace{5mm}

\noindent{\large \bf Acknowledgments}

SDO would like to thank the members of the Dept. ECM, Barcelona
University, for the kind hospitality.
This work has been partly supported by DGICYT (Spain) and by
CIRIT
(Generalitat de Catalunya).

\end{document}